\documentclass[copyright,creativecommons]{eptcs}
\providecommand{\event}{PLACE 2022} 

\usepackage{breakurl}             
\usepackage{underscore}           

\usepackage{booktabs}   
\usepackage{stmaryrd}
\usepackage{amsmath}
\usepackage{amssymb}
\usepackage{color}
\usepackage{listings}
\usepackage{mathpartir}
\usepackage{pifont}

\usepackage{tcolorbox}
\usepackage{tabularx}
\usepackage{xcolor}
\usepackage{wrapfig}
\usepackage{bm}

\usepackage{float}
\usepackage[sort]{cite}

\newcommand{\lns}{\textsc{Lang-n-Send}}
\newcommand{\channel}[1]{\textit{#1}}
\newcommand{\name}[1]{\textit{#1}}

\newcommand{\app}{\;}
\newcommand{\key}[1]{\ensuremath{\mathtt{#1}}}
\newcommand{\LNC}[1]{#1}

\newcommand{\labeledStep}[1]{\stackrel{#1}{\step}}

\newcommand{\trace}{\textit{tr}}
\newcommand{\traceBig}{\mathfrak{T}}
\newcommand{\langSet}{\textsc{lang-var}}
\newcommand{\traceSet}{\textsc{trace-var}}

\newcommand{\isInTrace}[4]{\key{isInTrace}(#1,#2) \Rightarrow #3 ~;~  #4}
\newcommand{\isInTraceOp}{\key{isInTrace}}

\newcommand{\stepL}{\longrightarrow_{\textsf{lan}}}
\newcommand{\stepExec}{\longrightarrow_{\textsf{exe}}}

\newcommand{\execLNN}[5]{\texttt{(}#1,#5\texttt{)}\texttt{>}_{#2}\app #4}

\newcommand{\LangDef}{\mathcal{L}}

\newcommand{\langVar}[1]{#1}

\newcommand{\unionLop}{\key{union}}
\newcommand{\ruletag}[1]{\textsc{(#1)}}
\newcommand{\subsnx}{_{\textsf{snx}}}
\newcommand{\unionLanguage}[2]{#1 \cup\subsnx #2}

\definecolor{magenta(dye)}{rgb}{0.79, 0.08, 0.48}
\definecolor{bondiblue}{rgb}{0.0, 0.58, 0.71}
\definecolor{navyblue}{rgb}{0.0, 0.0, 0.5}
\definecolor{lightskyblue}{rgb}{0.53, 0.81, 0.98}
\newcommand{\lop}[1]{#1}
\newcommand{\userLan}[1]{#1}

\newcommand*\colourcheck[1]{%
  \expandafter\newcommand\csname #1check\endcsname{\textcolor{#1}{\text{\ding{52}}}}%
}
\colourcheck{blue}
\colourcheck{green}
\colourcheck{red}

\usepackage{mathtools}
\usepackage{relsize}

\newcommand{\ninference}[3]{\inferrule[(#1)]{#2}{#3}}

\newcommand{\ba}{\begin{array}}
\newcommand{\ea}{\end{array}}

\newenvironment{syntax}{\[\ba{l@{\;\;}lcl}}{\ea\]}

\newcommand{\step}{\longrightarrow}

\definecolor{lightblue}{rgb}{0.25,0.25,1}

\definecolor{lightgray}{gray}{0.9}
\definecolor{darkergrey}{rgb}{0.75, 0.75, 0.75}

\usepackage{semantic}

\title{Lang-n-Send: Processes That Send Languages}
\author{Matteo Cimini
\institute{University of Massachusetts Lowell \\ Lowell, MA, USA}
\email{matteo\_cimini@uml.edu}
}
\def\titlerunning{Lang-n-Send: Processes That Send Languages}
\def\authorrunning{M. Cimini}
\begin{document}
\maketitle

\begin{abstract}
We present $\lns$, a $\pi$-calculus that is equipped with language definitions. 
Processes can define languages in operational semantics, and use them to execute programs. 
Furthermore, processes can send and receive pieces of operational semantics through channels.  

We present a reduction semantics for $\lns$, 
and we offer examples that demonstrate some of the scenarios that $\lns$ captures. 
\end{abstract}

\section{Introduction}


In the last decades, we have seen significant advances in language semantics tools that make it possible for programmers to quickly define and deploy their own programming languages and domain-specific languages, and use them in their programming solutions  \cite{LangWorkbenches}.

%

It is not too far in the future that it would just be the common practice for programmers to upload, in servers like Amazon AWS,
code that does not belong to a programming language that has been fixed beforehand, but rather belongs to a language that has been created on the fly. 
Programmers would upload both the program and the language in which the program must be evaluated. 
Reuse is fundamental in this scenario. Libraries of programming languages constructs, as envisioned by Peter Mosses' Component-Based Semantics \cite{Mosses08}, for example, can become the norm. 
Servers can provide pieces of languages to clients, which can use them to complete their own language and, in turn, send the language so built to (computing) servers to execute programs. 

Current literature does not offer a foundation that directly formalizes this and similar scenarios. 
In this paper, we present our work towards such a formal foundation. 

We present $\lns$, a $\pi$-calculus that is equipped with language definitions. 
Processes can define languages in operational semantics, add pieces of operational semantics together, and use them to execute programs. 
Processes can also send and receive pieces of languages through channels. 
After executing programs, $\lns$ processes can also send the trace of executions to other processes, which in turn can analyze these traces. 

We present a reduction semantics for $\lns$, and we provide some selected examples that demonstrate the scenarios that $\lns$ captures. 
We have specifically chosen examples that involve the communication of languages among processes. 
We show the following examples: 
\begin{itemize}
\item A client that, when entering a sensitive region of code, asks a server to provide the semantics of an interrupt operator, adds it to its language, and only then executes the code. 
\item A client that defines a language with an interrupt operator, but lets a server decide the semantics of the interruption (whether interrupt or disrupt semantics) by receiving, from the server, the rest of the rules that complete the semantics of the operator. 
\item  A client that lets a server decide whether its language is synchronous or asynchronous by receiving the semantics of the output operator from the server. 
\end{itemize}

We believe that $\lns$ represents a first step towards a firm foundation for this type of programming. 
The next section presents the syntax that $\lns$ uses to define languages. 
Section \ref{lns} presents the syntax of $\lns$ processes. 
Section \ref{operationalSemantics} presents a reduction semantics.  
Section \ref{examples} provides examples. 
Section \ref{relatedWork} discusses related work, and 
Section \ref{conclusion} concludes the paper.

\section{Syntax for Language Definitions}\label{syntaxLanguage}

The syntax of $\lns$ consists of two parts: 
the syntax for creating languages, 
and a $\pi$-calculus with language definitions. 
Language definitions can be created in operational semantics. 
The syntax that we adopt is inspired by \cite{lnp}, and is the following, where 
$\name{cname} \in \textsc{CatName}$, $\userLan{\langVar{X}} \in \textsc{Meta-Var}$, 
  $\name{opname} \in \textsc{OpName}$, and $\name{pn} \in \textsc{PredName}$. 
\begin{syntax}
   \text{\sf Language} & \userLan{\LangDef} & ::= & \userLan{(G,I)} \\
   \text{\sf Grammar} & \userLan{G} & ::= & \userLan{ s_1 \app \cdots \app s_n } \\
   \text{\sf {Grammar Rule}} & \userLan{s} & ::= & \name{cname} \app \langVar{X} ::= t_1\app  \mid \cdots \app \mid \app  t_n\\
   \text{\sf Inference System} & \userLan{I} & ::= & \userLan{ r_1 \app \cdots \app r_n }\\
   \text{\sf Rule} & \userLan{r} & ::= & \userLan{\inference{f_1 \app \cdots \app f_n }{f}}\\
   \text{\sf Formula} & \userLan{f} & ::= & (\name{pn} \app t_1 \cdots\app  t_n)  \\
   \text{\sf Term} & \userLan{t} & ::= &  \userLan{\langVar{X}} \mid \userLan{(\name{opname} \app t_1 \cdots\app  t_n)} 
\end{syntax}

\textsc{CatName} is a set of grammar category names such as \name{Process}, and \name{Action}. 
\textsc{Meta-Var} is a set of meta-variables.  
\textsc{OpName} is a set of constructor names such as \name{par} (for the parallel operator $\mid$), and \name{choice} (for the choice operator $+$). 
 \textsc{PredName} is a set of predicate names such as \name{step} (for reduction rules). 
 As names do not need to be strings, we shall use symbols for constructor and predicate names. 

A language has a grammar and an inference rule system. 
A grammar has multiple grammar rules, each of which defines a category name, and its meta-variable, by providing a series of grammar productions, which are terms.  
Terms are in abstract syntax tree style, that is, they have a top level constructor applied to a list of terms. 
We can demonstrate $\lns$ with languages that do not use binders. Therefore, we do not include syntax for binding, though it could be added.

An inference rule system has multiple rules, each of which has a series of formulae as premises, and a formula as conclusion. 
Formulae, too, are in abstract syntax tree style. 
Given a language definition, $\lns$ needs to invoke its evaluator to execute programs. 
As we need a way to locate such evaluator, we fix the following convention: 
The labeled transition relation is always $\step$, and its first argument is always the label of the transition, which is a term. (If reductions do not have labels, they would still use the first argument with a term that is never used). 

To make an example, let us consider Basic Process Algebra (BPA \cite{BergstraK84}) in its finite fragment (no recursion, nor definitions). 
BPA is formed with actions, the choice operator, and sequential composition. Below are the rules of BPA, where $a$ ranges over actions. 
Besides transitions of the form $P \labeledStep{a} P'$, BPA makes use of a predicate $P \step^{a}\checkmark$ that says that $P$ takes action $a$ and successfully terminates.

\begin{gather*}
a\step^{a}\checkmark
\quad
\inference{P_1\step^{a}\checkmark}{P_1+ P_2\step^{a}\checkmark} 
\quad
\inference{P_2\step^{a}\checkmark}{P_1+ P_2\step^{a}\checkmark} 
 \\[1ex]
\inference{P_1 \labeledStep{a} P_1'}{P_1+ P_2\labeledStep{a} P_1'} 
\quad
\inference{P_2 \labeledStep{a} P_2'}{P_1+ P_2\labeledStep{a} P_2'} 
\quad
\inference{P_1 \labeledStep{a} P_1'}{P_1\cdot P_2\labeledStep{a} P_1'\cdot P_2} 
\quad
\inference{P_1\step^{a}\checkmark}{P_1\cdot P_2\labeledStep{a}P_2} 
\end{gather*}

$\lns$ accommodates BPA as follows. 
The transition $P \labeledStep{a} P'$ is encoded as $(\step\app a\app P\app P')$. 
We represent the formula $P \step^{a}\checkmark$ with $(\name{checkMark} \app (a) \app P)$. Below, \name{act} is the operator for actions, and \name{seq} is the sequential operator.
We give this language the name \name{bpa}. 
\begin{align*}
\name{bpa} \triangleq (&Action \app A ::= (\name{a}) \mid (\name{b}) \mid (\name{c}) \quad\qquad \textit{(* We assume that the set of actions is $\{a,b,c\}$ *)} \\
&Process \app P ::=  (\name{act}\app A) \mid (\name{+}\app  P\app P)  \mid (\name{seq}\app  P\app P), 
\end{align*}
\begin{gather*}
(\name{checkMark}\app A \app  (\name{act}\app A))
\quad
\inference{(\name{checkMark}\app A \app P_1)}{(\name{checkMark}\app A\app (+\app P_1\app P_2))} 
\quad
\inference{(\name{checkMark}\app A \app P_2)}{(\name{checkMark}\app A\app (+\app P_1\app P_2))} 
 \\[1ex]
\inference{{(\step\app A \app P_1 \app P_1')}}{(\step\app A\app (+\app P_1\app P_2)\app P_1')} 
\quad
\inference{{(\step\app A \app P_2 \app P_2')}}{(\step\app A\app (+\app P_1\app P_2)\app P_2')} 
 \\[1ex]
\inference{{(\step\app A \app P_1 \app P_1')}}{(\step\app A\app (\name{seq}\app P_1\app P_2)\app (\name{seq}\app P_1'\app P_2))}
\quad
\inference{(\name{checkMark}\app A \app P_1)}{(\step\app A\app (\name{seq}\app P_1\app P_2)\app  P_2)} ~~)
\end{gather*}

\section{Syntax for Processes}\label{lns}

The syntax of $\lns$ processes $P$, $Q$, $R$, and so on, is defined as follows.
%
\begin{syntax}
   \text{\sf Trace} & \traceBig & ::= &  \trace \mid t \app \cdots \app t \\
   \text{\sf Language Builder} & \lop{\ell} & ::= &  l \mid \userLan{\mathcal{L}} \mid \lop{\ell \app \unionLop \app \ell} \\
   \text{\sf Processes} & P & ::= &\mathbf{0} \mid ~ x(y).P ~\mid~ \overline{x} \langle y\rangle.P  \mid~ P \mid P ~\mid~ P + P ~\\
   &&&\mid~ (\nu x).P ~\mid~ !P \\
  \textit(using ~languages)       &&&    
   		\mid \execLNN{\ell}{x}{pn_2}{t}{\traceBig} 
		\\
  \textit(analyzing ~executions)       &&&  \mid \isInTrace{t}{\traceBig}{P}{P} \\ 
  \textit(communicating ~languages) 		&&& \mid ~ x(l).P ~\mid~ \overline{x} \langle \ell \rangle.P \\ 
  \textit(communicating ~traces) 		&&& \mid ~ x(\trace).P ~\mid~ \overline{x} \langle \traceBig \rangle.P \\ 
\end{syntax}

$\lns$ contains the elements of the $\pi$-calculus \cite{MILNER19921,MILNER199241}. Channels are $x$, $y$, $z$, and so on. 
We assume a set $\langSet$ of variables $l$ for languages, and a set $\traceSet$ of variables $\trace$ for traces. 
These sets are distinct from each other, and from the set of channels. 

\emph{Language builder expressions} $\ell$ build a language $\userLan{\mathcal{L}}$, i.e., they ultimately evaluate to a language $\userLan{\mathcal{L}}$. 
This category can be extended with sophisticated language manipulations. 
We keep our syntax with the minimal set of operations that are enough to demonstrate our approach. 
Thus, we have included only a union operation for languages. 
$\unionLop$ adds new grammar productions and inference rules to a language. 
For example, $\name{bpa} \app \unionLop \app (Process \app P ::= (\name{loopOnNil}\app  P) ) ~ (\step\app A \app (\name{loopOnNil}\app (\name{nil}))  \app (\name{loopOnNil}\app (\name{nil})) )\,)$ returns the language with the same grammar productions for \name{Action}, and with the additional production $(\name{loopOnNil}\app  P)$ for $P$.  
Also, the rule above for \name{loopOnNil} is added to the rules of \name{bpa}.

$\lns$ extends the processes of the $\pi$-calculus with the following constructors. 
A process $\execLNN{\userLan{\lop{\ell}}}{x}{\userLan{pn_2}}{\userLan{{t}}}{\traceBig}$ is a \emph{program execution}. 
This process executes the program $t$ according to the operational semantics defined in the language $\ell$. 
In particular, we query the language for reductions that are provable with $\step$. 
We also keep track of the trace of executions. 
Traces are sequences of labels.  
As we use terms to represent labels, our traces $\traceBig$ are sequences of terms.
We assume that a program execution starts with an empty sequence of labels, 
which we denote with $[]$ to avoid a confusing empty space in our examples. 
A reduction step of $t$ carries a label, and we append it to $\traceBig$. 
Then, $\traceBig$ contains all the labels of all the steps of $t$, that is, $\traceBig$ is a trace of the execution of $t$. 
When the execution of $t$ terminates, the trace is sent over the channel $x$. 

To make an example: 

$\execLNN{\name{bpa}}{x}{\userLan{pn_2}}{
(\name{seq}\app (\name{act}\app (a))\app (\name{seq}\app (\name{act}\app (b))\app (\name{act}\app (c))))
}{[]}$ 
reduces to 

$\execLNN{\name{bpa}}{x}{\userLan{pn_2}}{
(\name{seq}\app (\name{act}\app (b))\app (\name{act}\app (c)))
}{(a)}$ 
which reduces to 

$\execLNN{\name{bpa}}{x}{\userLan{pn_2}}{
(\name{act}\app (c))
}{(a) ~ (b)}$ 
which reduces to 
$\overline{x} \langle (a) ~ (b) \rangle.\mathbf{0}$. 

Notice that, in BPA, this last $c$ does not take a transition, but $c \step^{c}\checkmark$ holds.
We could account for this with a straightforward modification of BPA that performs the last action as a labeled transition, but we prefer to use its original formulation. 

A process $\isInTrace{t}{\traceBig}{P}{Q}$ checks whether the label $t$ is one of the labels in the trace $\traceBig$. In that case, this process behaves as $P$, 
otherwise it behaves as $Q$. 

A process $x(l).P$ is a \emph{language input prefix}. This process receives a language on the channel $x$, binds it to $l$, and continues as $P$. 
A process $\overline{x} \langle \ell \rangle.P$ is a \emph{language output prefix}. This process evaluates $\ell$ to a language $\userLan{\mathcal{L}}$, sends it over the channel $x$, and continues as $P$. Similarly, a process $x(\trace).P$ is a \emph{trace input prefix} and receives traces. A process $\overline{x} \langle \traceBig \rangle.P$ is a \emph{trace output prefix} and sends traces.

\section{A Reduction Semantics for $\lns$}\label{operationalSemantics}

Figure \ref{fig:dynamicsemantics} shows the reduction semantics of $\lns$. 
Structural congruence $\equiv$ is standard. 
The reduction relation for the processes of $\lns$ is $\step$. This relation relies on two auxiliary relations: the \emph{language building reduction relation} $\stepL$, 
and the \emph{program reduction relation} $\stepExec$. Below we describe the rules of Figure \ref{fig:dynamicsemantics}. 

Rule \ruletag{comm} is standard. Rule \ruletag{comm-lang} handles the communication of languages. 
In this rule, $\stepL^{*}$ is the reflexive and transitive closure of $\stepL$. 
We evaluate $\ell$ to a language $\userLan{\mathcal{L}}$, and only then we perform the passing. 
Rule \ruletag{comm-trace} handles the communication of traces. 
Substitution $P\{\userLan{\mathcal{L}}/l\}$ substitutes the free occurrences of $l$ in $P$ with $\userLan{\mathcal{L}}$. 
Substitution $P\{\userLan{\traceBig}/\trace\}$ substitutes the free occurrences of $\trace$ in $P$ with $\traceBig$. 
Both substitutions are capture-avoiding, and their definition is straightforward, so we do not show it. 

Rule \ruletag{exec} handles program executions when the language is available, that is, it has been evaluated to some $\userLan{\mathcal{L}}$. 
This rule simply relies on $\stepExec$. Rule \ruletag{exec-ctx} evaluates $\ell$ with $\stepL$-reductions. 

Rules \ruletag{is-in-trace1} and \ruletag{is-in-trace2} define the behavior of $\isInTraceOp$. 
This process takes a step to $P$ if the label is in $\traceBig$, and takes a step to $Q$ otherwise. 

Rule \ruletag{union} performs the union of two languages using the operation $\cup\subsnx$. This operation adds new grammar productions and inference rules to a language in the way that we have seen. This operation has been previously defined in \cite{lnp}. (We discuss related work in Section \ref{relatedWork}.) Rules \ruletag{union-ctx1} and \ruletag{union-ctx2} evaluate the first and second argument of \key{union}, respectively. 

Rule \ruletag{program-step} handles program executions $\execLNN{\userLan{\mathcal{L}}}{x}{\userLan{pn_2}}{\userLan{{t}}}{\traceBig}$. 
This rule is responsible for executing $t$ according to the operational semantics of $\userLan{\mathcal{L}}$. 
To do so, we should query the inference rule system in $\userLan{\mathcal{L}}$. 
However, $\userLan{\mathcal{L}}$ contains \emph{syntax that represents} an inference system. 
We adopt the solution used in \cite{lnp}: we translate the language into a higher-order logic program with $\llbracket {\userLan{\mathcal{L}}} \rrbracket^{\textsf{lp}}$, 
and we use the provability relation $\models$ of logic programs to check whether a step from $t$ is provable for some target $t'$ and some label $t''$.
The translation $\llbracket {\userLan{\mathcal{L}}} \rrbracket^{\textsf{lp}}$ to logic programs is easy, and has been described in \cite{lnp}. The way this translation works was not novel in there either, as it has been demonstrated previously that inference systems of the like map well into logic programs \cite{Pfenning1999,Miller:2012lp}.
The provability relation $\models$ comes directly from the semantics of higher order logic programs, which can be found in \cite{Miller:2012lp}. 
Rule \ruletag{program-step} also appends $t''$ to the trace recorded in the program execution. 

Rule \ruletag{program-end} detects that a step is not provable for $t$. 
Then, the execution of $t$ is terminated, and we send the trace over the channel $x$. 

Notice that $t$ may fail to prove a step for several reasons, including that $t$ is stuck because of missing reduction rules in an ill-defined language. 
Programmers are responsible for giving well-designed languages, as $\lns$ does not check that. 

\begin{figure}[htbp]
{\small
\textsf{Reduction Semantics}  \hfill \fbox{$P \equiv P$, \;$P \step P$, \;$\ell \stepL \ell$, \;$P \stepExec P$}
\begin{gather*}
P \mid \mathbf{0} \equiv P 
\qquad
P \mid Q \equiv Q \mid P
\qquad
(P \mid Q) \mid R \equiv P \mid (Q \mid R)  
\qquad
!P \equiv P \mid !P
\\[1ex]
 (\nu x).\mathbf{0} \equiv \mathbf{0}
\qquad
 (\nu x).(\nu y).P \equiv (\nu y).(\nu x).P
\qquad
 (\nu x).(P \mid Q) \equiv (\nu x).P \mid Q,
~
 \textit{if $x$ is not a free name of $Q$}
\\[2ex]
\inference
	{P_1\step P_1'}
	{P_1\mid P_2 \step P_1'\mid P_2}
	\qquad
\inference
	{P\step P'}
	{(\nu x).P \step  (\nu x).P'}
\qquad
\inference
	{P \equiv P' & P' \step Q' & Q' \equiv Q}
	{P \step Q}
\\[2ex]
\ninference{comm}{}
	{
	x(y).P \mid \overline{x} \langle z\rangle.Q  \step  P\{z/y\} \mid Q
	}
\qquad
\ninference{comm-lang}
{\ell \stepL^{*} \userLan{\mathcal{L}}}
	{
	x(l).P \mid \overline{x} \langle \ell\rangle.Q \step  P\{\userLan{\mathcal{L}}/l\} \mid Q
	}
\qquad
\ninference{comm-trace}
{}
	{
	x(\trace).P \mid \overline{x} \langle \traceBig\rangle.Q  \step  P\{\userLan{\traceBig}/\trace\} \mid Q
	}
\\[2ex]
\ninference{exec}
{
\execLNN{\userLan{\mathcal{L}}}{x}{\userLan{pn_2}}{\userLan{\LNC{t}}}{\traceBig} 
\stepExec
P}
{
\execLNN{\userLan{\mathcal{L}}}{x}{\userLan{pn_2}}{\userLan{\LNC{t}}}{\traceBig} 
\step
P
}
\qquad 
\ninference{exec-ctx}
{\ell \stepL \ell'}
{
\execLNN{\userLan{\lop{\ell}}}{x}{\userLan{pn_2}}{\userLan{\LNC{t}}}{\traceBig} 
\step
\execLNN{\userLan{\lop{\ell'}}}{x}{\userLan{pn_2}}{\userLan{\LNC{t}}}{\traceBig}
}
\\[2ex]
\ninference{is-in-trace1}
{\traceBig = t_1 \cdots t_n \\ 1\leq i\leq n}
{
\isInTrace{t_i}{\traceBig}{P}{Q}
\step
P
}
\qquad
\ninference{is-in-trace2}
{\traceBig = t_1 \cdots t_n \\ t \not= t_i, \textit{ for all }1\leq i\leq n}
{
\isInTrace{t}{\traceBig}{P}{Q}
\step
Q
}
\\[2ex]
\ninference{union}
{}
{\lop{\userLan{\mathcal{L}_1}\app \unionLop \app \userLan{\mathcal{L}_2}} \stepL \unionLanguage{\userLan{\mathcal{L}_1}}{\userLan{\mathcal{L}_2}} 
} 
\qquad
\ninference{union-ctx1}
{\ell_1 \stepL \ell_1'}
{{\ell_1}\app \unionLop \app{\ell_2} \stepL {\ell_1'}\app \unionLop \app{\ell_2} 
} 
\qquad
\ninference{union-ctx2}
{\ell_2 \stepL \ell_2'}
{{\ell_1}\app \unionLop \app{\ell_2} \stepL {\ell_1}\app \unionLop \app{\ell_2'} 
} 
\\[2ex]
\ninference{program-step}
{ \llbracket {\userLan{\mathcal{L}}} \rrbracket^{\textsf{lp}} \models  \userLan{(\LNC{t} \step^{t''} \LNC{t'})} 
} 
{
\execLNN{\userLan{\mathcal{L}}}{x}{\userLan{pn_2}}{\userLan{\LNC{t}}}{{\traceBig}}
\stepExec 
\execLNN{\userLan{\mathcal{L}}}{x}{\userLan{pn_2}}{\userLan{\LNC{t'}}}{{{\traceBig} \app t''}}
} 
\qquad
\ninference{program-end}
{
\llbracket {\userLan{\mathcal{L}}} \rrbracket^{\textsf{lp}} \not\models  \userLan{(t \step^{t''} t')}  
} 
{
\execLNN{\userLan{\mathcal{L}}}{x}{\userLan{pn_2}}{\userLan{{t}}}{\traceBig}
\stepExec 
\overline{x} \langle \traceBig\rangle.\mathbf{0}
} 
\end{gather*}
}
\caption{Reduction semantics of $\lns$.}
\label{fig:dynamicsemantics}
\end{figure}

\section{Examples}\label{examples}

\paragraph{Server Provides a Disrupt Operator}
In this example, \textit{server} is a server that offers two services: \textit{task} and \textit{quitOnFailureTask}.
These tasks are executed with BPA processes. 
However, \textit{quitOnFailureTask} is critical, and should stop if a mistake is detected. 
BPA does not have a way to model disruptions. 
Therefore, at the moment of executing \textit{quitOnFailureTask} (and only in that case), 
\textit{server} requests the piece of operational semantics of the disrupt operator of LOTOS \cite{ISO-LOTOS89}, adapted for BPA in \cite{Baeten00modetransfer}, and adds it to the language \name{bpa} from Section \ref{syntaxLanguage}. 
Intuitively, $P \blacktriangleright Q$ means that $P$ can be disrupted by $Q$. This process behaves as $P$, though at any point, non-deterministically, $Q$ can start its computation, which discards $P$ forever. We repeat the inference rules for $\blacktriangleright$ (\hspace{1sp}\cite{Baeten00modetransfer}). 
\begin{gather*}
\inference{P_1\step^{a}\checkmark}{P_1\blacktriangleright P_2\step^{a}\checkmark}
\quad 
\inference{P_1 \labeledStep{a} P_1'}{P_1\blacktriangleright P_2\labeledStep{a} P_1'\blacktriangleright P_2} 
\\
\inference{P_2 \labeledStep{a} P_2'}{P_1\blacktriangleright P_2\labeledStep{a} P_2'} 
\quad 
\inference{P_2\step^{a}\checkmark}{P_1\blacktriangleright P_2\step^{a}\checkmark}
\end{gather*}

We define the $\lns$ counterpart of $\blacktriangleright$ in two parts. 
\textit{almostDisrupt} contains the first row of the rules above. These rules define the behavior of $\blacktriangleright$ insofar the preempted process is concerned. 
\textit{disruptRules} contains the second row of rules, which are for the preempting process. 
Then, \textit{disrupt} contains the union of the two, and is the piece of language with the full definition of $\blacktriangleright$.\\ 

\textit{almostDisrupt} $\triangleq$ $(Process \app P ::=  (\blacktriangleright \app  P\app P), $\\[1ex]
\indent\indent\indent\indent\
\inference{{(\name{checkMark}\app A \app P_1)}}{(\name{checkMark}\app A\app (\blacktriangleright \app P_1\app P_2))} 
\quad 
\inference{{(\step\app A \app P_1 \app P_1')}}{(\step\app A\app (\blacktriangleright\app P_1\app P_2)\app (\blacktriangleright\app P_1'\app P_2))} ~~)

\textit{disruptRules} $\triangleq$
$(~~ 
\inference{{(\step\app A \app P_2 \app P_2')}}{(\step\app A\app (\blacktriangleright\app P_1\app P_2)\app P_2')} 
\quad 
\inference{{(\name{checkMark}\app A \app P_2)}}{(\name{checkMark}\app A\app (\blacktriangleright \app P_1\app P_2))} ~~)
$

$\textit{disrupt} \triangleq \textit{almostDisrupt} \app \unionLop \app \textit{disruptRule}$\\

Below, the process \textit{disruptOperatorProvider} is a server, different from \textit{server}, that provides the \textit{disrupt} piece of language over the channel \channel{getDisrupt}. The code for \textit{server} is also below. We assume that \textit{bpa\_program}, a term, is a BPA process to be executed for \channel{quitOnFailureTask}, and that \textit{bpa\_sorry} is the BPA process that can non-deterministically preempt \textit{bpa\_program}. For readability, we use $\blacktriangleright$ in infix notation. 
The process for \channel{task} is irrelevant, and we chose $(\name{act}\app (a))$.\\

\newenvironment{nospaceflalign*}
 {\setlength{\abovedisplayskip}{0pt}\setlength{\belowdisplayskip}{0pt}%
  \csname flalign*\endcsname}
 {\csname endflalign*\endcsname\ignorespacesafterend}

$\textit{disruptOperatorProvider} \triangleq !(\overline{\channel{getDisrupt}}\langle\textit{disrupt}\rangle)$

$
\begin{array}{ll} 
 \textit{server} \triangleq \app &!(~\channel{task}(x).\execLNN{\textit{bpa}}{x}{}{(\name{act}\app (a))}{[]} \\ 
 &~~+\\
 &~~ \channel{quitOnFailureTask}(x).\channel{getDisrupt}(l).\execLNN{\textit{bpa\app \unionLop \app l}}{x}{}{(\textit{bpa\_program} \blacktriangleright \textit{bpa\_sorry})}{[]}\;)
\end{array}
$

\indent$\textit{system} \triangleq (\textit{server}  \mid \textit{disruptOperatorProvider} \mid \textit{client}_1 \mid \textit{client}_2 \app \ldots \app \mid \textit{client}_n)$\\

Suppose that \textit{bpa\_sorry} performs the action (\textit{sorry}). 
We can detect whether \textit{bpa\_program} has been disrupted with $\isInTraceOp$. 
The second branch of the choice operator of \textit{server} would be\\

$
\nonumber \channel{quitOnFailureTask}(x).\channel{getDisrupt}(l).(\nu x).\\
\indent\indent\indent (\execLNN{\textit{bpa\app \unionLop \app l}}{x}{}{(\textit{bpa\_program} \blacktriangleright \textit{bpa\_sorry})}{[]}
									\mid 
							x(tr).\isInTrace{(\textit{sorry})}{tr}{P_1}{P_2})
$\\

Here, \textit{server} creates a private channel $x$ over which the trace is sent. 
We assume that $P_1$ and $P_2$ are two processes that \textit{server} cares to execute depending on whether (\textit{sorry}) is in the trace or not. 

\paragraph{Server Decides Disrupt vs Interrupt}

In this example, the server \textit{disruptOperatorProvider} is called \textit{quitModeProvider}. 
It takes in input a channel (such as \channel{quitOnFailureTask}), and non-deterministically decides whether to provide the disrupt operator or the interrupt operator $\rhd$ from \cite{Baeten00modetransfer}. The process $P \rhd Q$ means that $P$ can be interrupted by $Q$. Differently from the disrupt operator, which completely discards $P$ when $Q$ takes over, the interrupt operator resumes $P$ after $Q$ terminates. 

\textit{bpa\_program} uses \emph{one} operator whose underlying semantics is given by \textit{quitModeProvider}. 
We fix the symbol for this operator to be $\blacktriangleright$. 
Therefore, when \textit{quitModeProvider} gives the interrupt semantics, it does so by giving the rules of $\rhd$ for the symbol $\blacktriangleright$. 
The piece of language for the preempted process, \textit{almostDisrupt}, is the same for $\blacktriangleright$ and $\rhd$. 
The rules for the preempting process are the following (\hspace{1sp}\cite{Baeten00modetransfer}). 
\begin{gather*}
\inference{P_2 \labeledStep{a} P_2'}{P_1\rhd P_2\labeledStep{a} P_2' \cdot P_1} 
\quad 
\inference{P_2\step^{a}\checkmark}{P_1\rhd P_2\labeledStep{a} P_1}
\end{gather*}

Below, \textit{interruptRules} contains the $\lns$ counterpart of these rules, though defined for the symbol $\blacktriangleright$, as explained above.   
When we add \textit{interruptRules} to \textit{almostDisrupt} we obtain the full definition of the interrupt operator (given as $\blacktriangleright$), which we call \textit{interrupt}.

\textit{interruptRules} $\triangleq$ 
(~$
\inference{{(\step\app A \app P_2 \app P_2')}}{(\step\app A\app (\blacktriangleright\app P_1\app P_2)\app (\name{seq} \app P_2'\app P_1))} 
\quad 
\inference{{(\name{checkMark}\app A \app P_2)}}{(\step\app A\app (\blacktriangleright\app P_1\app P_2)\app  P_1)} 
$~)
\\[0.5ex]

$\textit{interrupt} \triangleq \textit{almostDisrupt} \app \unionLop \app \textit{interruptRules}$


$\textit{quitModeProvider} \triangleq~ !\channel{whatTask}(y).(~\overline{\channel{getQuitMode}}\langle\textit{interrupt}\rangle 
\;+\; 
\overline{\channel{getQuitMode}}\langle\textit{disrupt}\rangle~)
$

$
\begin{array}{ll} 
\nonumber \textit{server} \triangleq~&!( \channel{task}(x).\execLNN{\textit{bpa}}{x}{}{(\name{act}\app (a))}{[]} \\
\nonumber &~~+\\
\nonumber &~~ \channel{quitOnFailureTask}(x).\overline{\channel{whatTask}}\langle\textit{quitOnFailureTask}\rangle. \channel{getQuitMode}(l).\\
\nonumber &\qquad \qquad  \qquad \qquad \qquad\qquad\qquad  \execLNN{\textit{bpa\app \unionLop \app l}}{x}{}{(\textit{bpa\_program} \blacktriangleright \textit{bpa\_sorry})}{[]} ~)
\end{array}
$

$\textit{system} \triangleq (\textit{server}  \mid \textit{quitModeProvider} \mid \textit{client}_1 \mid \textit{client}_2 \app \ldots \app \mid \textit{client}_n)$

\paragraph{Server Decides Synchronous vs Asynchronous}

In this example, the process \textit{client} executes a CCS process called \textit{ccs\_program}. However, \textit{client} requests the semantics of the output prefix operator from the server \textit{outputProvider}, 
which decides, non-deterministically, whether \textit{ccs\_program} must be executed synchronously or asynchronously. 
There is a syntactic difference between the synchronous output $\overline{a}.P$ and the asynchronous output $\overline{a}$ (with no continuation process). 
As \textit{ccs\_program} is fixed, we settle to use the more general output form $\overline{a}.P$, though its semantics will be given by the server. 

We define a partial CCS with inaction, input prefix, output prefix, a one-channel restriction operator $P\backslash a$, and the parallel operator. 
The semantics of the output prefix, however, is not given. 
As we do not have negative premises in rules, we define $P\backslash a$ by cases. 
For simplicity, we only include channels $x$ and $y$. \\

$
\begin{array}{l}
\name{Channel} \app a ::= x \mid y \\
\name{Label} \app L ::= \tau \mid a \mid \overline{a} \\
\name{Process} \app P ::=  \mathbf{0} \mid a.P \mid \overline{a}.P \mid~ P \mid P ~\mid P\backslash a
 \\[0.5ex]
\end{array}
$

\begin{gather*}
a.P \labeledStep{a} P 
\quad 
\inference{P \labeledStep{\tau} P'}
{P\backslash a \labeledStep{\tau} P'\backslash a} 
 \\[1ex]
\inference{P \labeledStep{y} P'}
{P\backslash x \labeledStep{y} P'\backslash x} 
\quad
\inference{P \labeledStep{\overline{y}} P'}
{P\backslash x \labeledStep{\overline{y}} P'\backslash x} 
\quad
\inference{P \labeledStep{x} P'}
{P\backslash y \labeledStep{x} P'\backslash y} 
\quad
\inference{P \labeledStep{\overline{x}} P'}
{P\backslash y \labeledStep{\overline{x}} P'\backslash y} 
%
%
\end{gather*}
\begin{gather*}
\inference{P_1 \labeledStep{L} P_1'}{P_1\mid P_2\labeledStep{L} P_1'\mid P_2} 
\quad
\inference{P_2 \labeledStep{L} P_2'}{P_1\mid P_2\labeledStep{L} P_1\mid P_2'} 
\\[1ex]
\inference{P_1 \labeledStep{a} P_1' & P_2 \labeledStep{\overline{a}} P_2'}{P_1\mid P_2\labeledStep{\tau} P_1'\mid P_2'}
\quad
\inference{P_1 \labeledStep{\overline{a}} P_1' & P_2 \labeledStep{{a}} P_2'}{P_1\mid P_2\labeledStep{\tau} P_1'\mid P_2'}
\end{gather*}

Below, \textit{partialCCS} contains the $\lns$ counterpart of the partial CCS defined above. 
Inaction is called \name{nil}, input prefix is called \name{in}, output prefix is called \name{out}, the restriction operator is called \name{res}, and the parallel operator is called \name{par}. \\

$\begin{array}{ll} \textit{partialCCS} \triangleq \\
&(Channel \app a ::= (\name{x}) \mid (\name{y})  \\
& Label \app L ::= (tau) \mid (\name{in}\app a) \mid (\name{out}\app a) \\
& Process \app P ::=  (\name{nil}) \mid (\name{in}\app a \app P) \mid (\name{out}\app a \app P) \mid (\name{res}\app  a\app P) \mid (\name{par}\app  P\app P), 
\end{array}
$
\begin{gather*}
{{(\step\app (\name{in}\app a) \app (\name{in}\app a\app P) \app P)}} 
\quad
\inference{{(\step\app (tau) \app P \app P')}}{(\step\app (tau)\app (\name{res}\app  a\app P)\app (\name{res}\app  a\app P'))} 
 \\[1ex]
\inference{{(\step\app (\name{in}\app (\name{y})) \app P \app P')}}{(\step\app (\name{in}\app (\name{y}))\app (\name{res}\app  (\name{x})\app P)\app (\name{res}\app  (\name{x})\app P'))} 
\quad
\inference{{(\step\app (\name{out}\app (\name{y})) \app P \app P')}}{(\step\app (\name{out}\app (\name{y}))\app (\name{res}\app  (\name{x})\app P)\app (\name{res}\app  (\name{x})\app P'))} 
\\[1ex]
\inference{{(\step\app (\name{in}\app (\name{x})) \app P \app P')}}{(\step\app (\name{in}\app (\name{x}))\app (\name{res}\app  (\name{y})\app P)\app (\name{res}\app  (\name{y})\app P'))} 
\quad
\inference{{(\step\app (\name{out}\app (\name{x})) \app P \app P')}}{(\step\app (\name{out}\app (\name{x}))\app (\name{res}\app  (\name{y})\app P)\app (\name{res}\app  (\name{y})\app P'))} 
\\[1ex]
\inference{{(\step\app L \app P_1 \app P_1')}}{(\step\app L\app (par\app P_1\app P_2)\app (par\app P_1'\app P_2))} 
\quad
\inference{{(\step\app L \app P_2 \app P_2')}}{(\step\app L\app (par\app P_1\app P_2)\app (par\app P_1\app P_2'))} 
\\[1ex]
\inference{{(\step\app (in\app A) \app P_1 \app P_1') & (\step\app (out\app A) \app P_2 \app P_2')}}{(\step\app (tau) \app (par\app P_1\app P_2)\app (par\app P_1'\app P_2'))} 
\quad
\inference{{(\step\app (out\app A) \app P_1 \app P_1') & (\step\app (in\app A) \app P_2 \app P_2')}}{(\step\app (tau) \app (par\app P_1\app P_2)\app (par\app P_1'\app P_2'))} ~~)
\end{gather*}

To complete \noindent\textit{partialCCS} with synchronous output, we add the usual rule for output prefix. 
To complete \noindent\textit{partialCCS} with asynchronous output, we add 
1) the asynchronous output $\overline{a}$ to the grammar, added as $(\name{out}'\app a)$ below, 2) its reduction rule $\overline{a} \labeledStep{\overline{a}} \mathbf{0}$, 
and 3) the rule $\overline{a}.P \labeledStep{\tau}(\overline{a} \mid P)$\footnote{Notice that a simple rule like $\overline{a}.P \labeledStep{\tau} \overline{a}.\mathbf{0} \mid P$ is problematic, as the rule applies to $\overline{a}.\mathbf{0}$, as well, replicating forever. 
Also notice that this $\tau$-transition does not resolve a choice, as \textit{partialCCS} does not contain $+$.
}:

$\textit{synchOutput} \triangleq (~ (\step\app (\name{out}\app a) \app (\name{out}\app a\app P) \app P)~)$
\qquad \textit{(* this rule is $\overline{a}.P \labeledStep{a} P$ *)}

$\textit{asynchOutput} \triangleq 
(~Process \app P ::=  (\name{out}'\app a), 
\\ 
\indent\indent\indent\indent\indent~~ (\step\app (\name{out}\app a) \app (\name{out}\textit{$'$}\app a\app) \app (\name{nil}))
\qquad
(\step\app (\name{tau}) \app (\name{out}\app a\app P) \app (\name{par}\app (\name{out}\textit{$'$}\app a)\app P))
~).
$

We give the definitions of \textit{outputProvider}, \textit{client}, and \textit{system} below. 
When \textit{ccs\_program} is the process $(\overline{x}.y.\mathbf{0} \mid \overline{y}.\mathbf{0}) \backslash x$, 
whether a communication over the channel $y$ happens or not depends on 
whether \textit{outputProvider} sends \textit{synchOutput} or \textit{asynchOutput}.

$\textit{outputProvider} \triangleq !(~\,\overline{\channel{getOutput}}\langle\textit{synchOutput}\rangle 
\;+\; 
\overline{\channel{getOutput}}\langle\textit{asynchOutput}\rangle~)
$

$\textit{client} \triangleq \channel{getOutput}(l).
		\execLNN{\textit{partialCCS} \app \unionLop \app l}{x}{}{{\textit{ccs\_program}}}{[]} 
$

$\textit{system} \triangleq \textit{client} \mid \textit{outputProvider}$

\section{Related Work}\label{relatedWork}

Our closest related work is \cite{lnp}. 
Such work offers a $\lambda$-calculus with first-class languages. 
We would like to characterize precisely the differences between this paper and that work. 
This paper embeds language definitions in the context of the $\pi$-calculus rather than the $\lambda$-calculus. 
The syntax for languages, the language union operator, and the translation to logic programs are from \cite{lnp}. 
The operator for program executions, and rule \textsc{(program-step)} are inspired by \cite{lnp}, 
but there are several differences in that \cite{lnp} does not allow for labeled transitions, and does not keep track of the trace of the execution. 
Moreover, as a consequence of this latter remark, \cite{lnp} does not have operations such as \key{isInTrace}, nor any other operation for analyzing executions. 
Furthermore, \cite{lnp} imposes that languages have a notion of values, and successful program executions terminate ending up with a value. 
This hardly applies to process algebras. 
\cite{lnp} does not make any example of concurrent scenarios. 
All the examples in this paper are new. 

Semantics engineering tools allow programmers to define their own programming languages \cite{redex,Rosu2010,Sewell:2007}. 
Language workbenches \cite{LangWorkbenches} go even further in that direction, 
and can automatically generate many components for the languages being defined, such as editors with syntax
colouring, highlighting, completion, and reference resolution, and they assist in code generation, as well as other phases. 
However, we are not aware of systems that allow pieces of languages to be sent and received. 

Multi-language operational semantics has been studied in several works. 
Matthews and Findler provide a seminal work of this field \cite{Matthews:2007zr}. Recent works in multi-language semantics are \cite{pattersonLink,funTal,fab}. 
All these works apply to two languages selected beforehand, and do not handle arbitrary languages specified by users. 
Furthermore, they do not offer a formal semantics of processes that communicate languages.

\section{Conclusion}\label{conclusion}

We have presented $\lns$, a $\pi$-calculus that is equipped with language definitions. 
Processes can define languages, and use them to execute programs. 
Moreover, processes can send and receive pieces of languages through channels.  
We have presented a reduction semantics for $\lns$. 

We have offered examples that show that $\lns$ can express concurrent scenarios that are not typical, where processes add language features based on semantics sent by servers, and where they obtain which semantics their operators adopt from servers.  
We believe that $\lns$ represents a first step to a firm foundation for this type of programming. 

In the future, we would like to extend $\lns$. 
Indeed, we see $\lns$ as a minimal foundational calculus that accommodates the communication of languages. 
We purposely did not include operations that, in fact, are interesting in this context. 
We plan to extend $\lns$ with more operations on languages, such as removing rules, and renaming operators, as well as more complex features such as converting languages from substitution-based to environment-based, among others. 

Adding binders to our language definitions does not seem to be problematic. \cite{lnp} has made that addition to model a $\lambda$-calculus as language definition. 
We plan to use binders to make examples with the $\pi$-calculus and its variants as $\lns$ language definitions. 

We plan to add more operations that query traces more precisely, such as counting labels in traces, and checking whether some labels appear before others. 
We also plan to add primitive operations for slicing the traces received \cite{slicing}, and we plan to add monitors to program executions \cite{monitorSurvey}.  

$\lns$ does not allow for the term of a terminated program execution to be sent. We have not included this feature because we believe that it enables rather complex dynamics, and we wanted to confine our examples to the already interesting scenarios that sending/receiving languages allow. 
We plan to explore the sending of terms after execution as future work. 

We plan to study more examples such as servers that decide the semantics of the parallel operator for client processes (CCS style, only interleaving and no communication, or the synchronous CSP parallel composition \cite{hoare1985communicating}, for instance). Another example is that of servers that decide the semantics of the choice operator, such as internal vs external, among other possibilities. 

Finally, we would like to study an appropriate notion of bisimilarity equivalence in this context.

\bibliographystyle{eptcs}
\bibliography{alllns}

\begin{thebibliography}{10}
\providecommand{\bibitemdeclare}[2]{}
\providecommand{\surnamestart}{}
\providecommand{\surnameend}{}
\providecommand{\urlprefix}{Available at }
\providecommand{\url}[1]{\texttt{#1}}
\providecommand{\href}[2]{\texttt{#2}}
\providecommand{\urlalt}[2]{\href{#1}{#2}}
\providecommand{\doi}[1]{doi:\urlalt{http://dx.doi.org/#1}{#1}}
\providecommand{\eprint}[1]{arXiv:\urlalt{https://arxiv.org/abs/#1}{#1}}
\providecommand{\bibinfo}[2]{#2}

\bibitemdeclare{inproceedings}{slicing}
\bibitem{slicing}
\bibinfo{author}{Hiralal \surnamestart Agrawal\surnameend} \&
  \bibinfo{author}{Joseph~R. \surnamestart Horgan\surnameend}
  (\bibinfo{year}{1990}): \emph{\bibinfo{title}{Dynamic Program Slicing}}.
\newblock In: {\sl \bibinfo{booktitle}{Proceedings of the ACM SIGPLAN 1990
  Conference on Programming Language Design and Implementation}},
  \bibinfo{series}{PLDI '90}, \bibinfo{publisher}{Association for Computing
  Machinery}, \bibinfo{address}{New York, NY, USA}, pp.
  \bibinfo{pages}{246--256}, \doi{10.1145/93542.93576}.

\bibitemdeclare{book}{Baeten00modetransfer}
\bibitem{Baeten00modetransfer}
\bibinfo{author}{Jos C.~M. \surnamestart Baeten\surnameend} \&
  \bibinfo{author}{Jan~A. \surnamestart Bergstra\surnameend}
  (\bibinfo{year}{2000}): \emph{\bibinfo{title}{Mode transfer in process
  algebra}}.
\newblock {\sl \bibinfo{series}{Computing Science Reports}}
  \bibinfo{volume}{00-01}, \bibinfo{publisher}{Technische Universiteit
  Eindhoven}.

\bibitemdeclare{article}{BergstraK84}
\bibitem{BergstraK84}
\bibinfo{author}{Jan~A. \surnamestart Bergstra\surnameend} \&
  \bibinfo{author}{Jan~W. \surnamestart Klop\surnameend}
  (\bibinfo{year}{1984}): \emph{\bibinfo{title}{Process Algebra for Synchronous
  Communication}}.
\newblock {\sl \bibinfo{journal}{Information and Control}}
  \bibinfo{volume}{60}(\bibinfo{number}{1-3}), pp. \bibinfo{pages}{109--137},
  \doi{10.1016/S0019-9958(84)80025-X}.

\bibitemdeclare{inproceedings}{monitorSurvey}
\bibitem{monitorSurvey}
\bibinfo{author}{Ian \surnamestart Cassar\surnameend}, \bibinfo{author}{Adrian
  \surnamestart Francalanza\surnameend}, \bibinfo{author}{Luca \surnamestart
  Aceto\surnameend} \& \bibinfo{author}{Anna \surnamestart
  Ing{\'{o}}lfsd{\'{o}}ttir\surnameend} (\bibinfo{year}{2017}):
  \emph{\bibinfo{title}{A Survey of Runtime Monitoring Instrumentation
  Techniques}}.
\newblock In: {\sl \bibinfo{booktitle}{Proceedings Second International
  Workshop on Pre- and Post-Deployment Verification Techniques, PrePost@iFM
  2017, Torino, Italy, 19 September 2017}}, pp. \bibinfo{pages}{15--28},
  \doi{10.4204/EPTCS.254.2}.

\bibitemdeclare{inproceedings}{lnp}
\bibitem{lnp}
\bibinfo{author}{Matteo \surnamestart Cimini\surnameend}
  (\bibinfo{year}{2021}): \emph{\bibinfo{title}{A Calculus for Multi-language
  Operational Semantics}}.
\newblock In: {\sl \bibinfo{booktitle}{Software Verification - 13th
  International Conference, {VSTTE} 2021, New Haven, CT, USA, October 18-19,
  2021, and 14th International Workshop, {NSV} 2021, Los Angeles, CA, USA, July
  18-19, 2021, Revised Selected Papers}}, pp. \bibinfo{pages}{25--42},
  \doi{10.1007/978-3-030-95561-8\_3}.

\bibitemdeclare{incollection}{LangWorkbenches}
\bibitem{LangWorkbenches}
\bibinfo{author}{Sebastian \surnamestart Erdweg\surnameend},
  \bibinfo{author}{Tijs \surnamestart Storm\surnameend},
  \bibinfo{author}{Markus \surnamestart Völter\surnameend},
  \bibinfo{author}{Meinte \surnamestart Boersma\surnameend},
  \bibinfo{author}{Remi \surnamestart Bosman\surnameend},
  \bibinfo{author}{WilliamR. \surnamestart Cook\surnameend},
  \bibinfo{author}{Albert \surnamestart Gerritsen\surnameend},
  \bibinfo{author}{Angelo \surnamestart Hulshout\surnameend},
  \bibinfo{author}{Steven \surnamestart Kelly\surnameend},
  \bibinfo{author}{Alex \surnamestart Loh\surnameend},
  \bibinfo{author}{Gabriël D.~P. \surnamestart Konat\surnameend},
  \bibinfo{author}{PedroJ. \surnamestart Molina\surnameend},
  \bibinfo{author}{Martin \surnamestart Palatnik\surnameend},
  \bibinfo{author}{Risto \surnamestart Pohjonen\surnameend},
  \bibinfo{author}{Eugen \surnamestart Schindler\surnameend},
  \bibinfo{author}{Klemens \surnamestart Schindler\surnameend},
  \bibinfo{author}{Riccardo \surnamestart Solmi\surnameend},
  \bibinfo{author}{Vlad~A. \surnamestart Vergu\surnameend},
  \bibinfo{author}{Eelco \surnamestart Visser\surnameend},
  \bibinfo{author}{Kevin \surnamestart Vlist\surnameend},
  \bibinfo{author}{Guido~H. \surnamestart Wachsmuth\surnameend} \&
  \bibinfo{author}{Jimi \surnamestart Woning\surnameend}
  (\bibinfo{year}{2013}): \emph{\bibinfo{title}{{The State of the Art in
  Language Workbenches}}}.
\newblock In \bibinfo{editor}{Martin \surnamestart Erwig\surnameend},
  \bibinfo{editor}{Richard~F. \surnamestart Paige\surnameend} \&
  \bibinfo{editor}{Eric \surnamestart Wyk\surnameend}, editors: {\sl
  \bibinfo{booktitle}{Software Language Engineering}}, {\sl
  \bibinfo{series}{Lecture Notes in Computer Science}} \bibinfo{volume}{8225},
  \bibinfo{publisher}{Springer}, pp. \bibinfo{pages}{197--217},
  \doi{10.1007/978-3-319-02654-1_11}.

\bibitemdeclare{book}{redex}
\bibitem{redex}
\bibinfo{author}{Matthias \surnamestart Felleisen\surnameend},
  \bibinfo{author}{Robert~Bruce \surnamestart Findler\surnameend} \&
  \bibinfo{author}{Matthew \surnamestart Flatt\surnameend}
  (\bibinfo{year}{2009}): \emph{\bibinfo{title}{Semantics Engineering with PLT
  Redex}}, \bibinfo{edition}{1st} edition.
\newblock \bibinfo{publisher}{The MIT Press}.

\bibitemdeclare{book}{hoare1985communicating}
\bibitem{hoare1985communicating}
\bibinfo{author}{C.A.R. \surnamestart Hoare\surnameend} (\bibinfo{year}{1985}):
  \emph{\bibinfo{title}{Communicating Sequential Processes}}.
\newblock \bibinfo{series}{Prentice-Hall International Series in Computer
  Science}, \bibinfo{publisher}{Prentice Hall}.
\newblock \urlprefix\url{http://www.usingcsp.com/cspbook.pdf}.

\bibitemdeclare{misc}{ISO-LOTOS89}
\bibitem{ISO-LOTOS89}
\bibinfo{author}{\surnamestart ISO/IEC\surnameend} (\bibinfo{year}{1989}):
  \emph{\bibinfo{title}{LOTOS --- a formal description technique based on the
  temporal ordering of observational behaviour}}.
\newblock \bibinfo{howpublished}{ISO IS 8807}, \doi{10.3403/00230466U}.

\bibitemdeclare{inproceedings}{Matthews:2007zr}
\bibitem{Matthews:2007zr}
\bibinfo{author}{Jacob \surnamestart Matthews\surnameend} \&
  \bibinfo{author}{Robert~Bruce \surnamestart Findler\surnameend}
  (\bibinfo{year}{2007}): \emph{\bibinfo{title}{Operational Semantics for
  Multi-Language Programs}}.
\newblock In: {\sl \bibinfo{booktitle}{Proceedings of the 34th Annual ACM
  SIGPLAN-SIGACT Symposium on Principles of Programming Languages}},
  \bibinfo{series}{POPL '07}, \bibinfo{publisher}{Association for Computing
  Machinery}, \bibinfo{address}{New York, NY, USA}, p. \bibinfo{pages}{3–10},
  \doi{10.1145/1190216.1190220}.

\bibitemdeclare{book}{Miller:2012lp}
\bibitem{Miller:2012lp}
\bibinfo{author}{Dale \surnamestart Miller\surnameend} \&
  \bibinfo{author}{Gopalan \surnamestart Nadathur\surnameend}
  (\bibinfo{year}{2012}): \emph{\bibinfo{title}{Programming with Higher-Order
  Logic}}, \bibinfo{edition}{1st} edition.
\newblock \bibinfo{publisher}{Cambridge University Press},
  \bibinfo{address}{New York, NY, USA}, \doi{10.1017/CBO9781139021326}.

\bibitemdeclare{article}{MILNER19921}
\bibitem{MILNER19921}
\bibinfo{author}{Robin \surnamestart Milner\surnameend},
  \bibinfo{author}{Joachim \surnamestart Parrow\surnameend} \&
  \bibinfo{author}{David \surnamestart Walker\surnameend}
  (\bibinfo{year}{1992}): \emph{\bibinfo{title}{A calculus of mobile processes,
  I}}.
\newblock {\sl \bibinfo{journal}{Information and Computation}}
  \bibinfo{volume}{100}(\bibinfo{number}{1}), pp. \bibinfo{pages}{1--40},
  \doi{10.1016/0890-5401(92)90008-4}.

\bibitemdeclare{article}{MILNER199241}
\bibitem{MILNER199241}
\bibinfo{author}{Robin \surnamestart Milner\surnameend},
  \bibinfo{author}{Joachim \surnamestart Parrow\surnameend} \&
  \bibinfo{author}{David \surnamestart Walker\surnameend}
  (\bibinfo{year}{1992}): \emph{\bibinfo{title}{A calculus of mobile processes,
  II}}.
\newblock {\sl \bibinfo{journal}{Information and Computation}}
  \bibinfo{volume}{100}(\bibinfo{number}{1}), pp. \bibinfo{pages}{41--77},
  \doi{10.1016/0890-5401(92)90009-5}.

\bibitemdeclare{inproceedings}{Mosses08}
\bibitem{Mosses08}
\bibinfo{author}{Peter~D. \surnamestart Mosses\surnameend}
  (\bibinfo{year}{2008}): \emph{\bibinfo{title}{Component-Based Description of
  Programming Languages}}.
\newblock In: {\sl \bibinfo{booktitle}{Visions of Computer Science - {BCS}
  International Academic Conference, Imperial College, London, UK, 22-24
  September 2008}}, pp. \bibinfo{pages}{275--286},
  \doi{10.14236/ewic/VOCS2008.23}.

\bibitemdeclare{inproceedings}{pattersonLink}
\bibitem{pattersonLink}
\bibinfo{author}{Daniel \surnamestart Patterson\surnameend} \&
  \bibinfo{author}{Amal \surnamestart Ahmed\surnameend} (\bibinfo{year}{2017}):
  \emph{\bibinfo{title}{{Linking Types for Multi-Language Software: Have Your
  Cake and Eat It Too}}}.
\newblock In \bibinfo{editor}{Benjamin~S. \surnamestart Lerner\surnameend},
  \bibinfo{editor}{Rastislav \surnamestart Bod{\'i}k\surnameend} \&
  \bibinfo{editor}{Shriram \surnamestart Krishnamurthi\surnameend}, editors:
  {\sl \bibinfo{booktitle}{2nd Summit on Advances in Programming Languages
  (SNAPL 2017)}}, {\sl \bibinfo{series}{Leibniz International Proceedings in
  Informatics (LIPIcs)}}~\bibinfo{volume}{71}, \bibinfo{publisher}{Schloss
  Dagstuhl--Leibniz-Zentrum fuer Informatik}, \bibinfo{address}{Dagstuhl,
  Germany}, pp. \bibinfo{pages}{12:1--12:15},
  \doi{10.4230/LIPIcs.SNAPL.2017.12}.

\bibitemdeclare{inproceedings}{funTal}
\bibitem{funTal}
\bibinfo{author}{Daniel \surnamestart Patterson\surnameend},
  \bibinfo{author}{Jamie \surnamestart Perconti\surnameend},
  \bibinfo{author}{Christos \surnamestart Dimoulas\surnameend} \&
  \bibinfo{author}{Amal \surnamestart Ahmed\surnameend} (\bibinfo{year}{2017}):
  \emph{\bibinfo{title}{FunTAL: Reasonably Mixing a Functional Language with
  Assembly}}.
\newblock In: {\sl \bibinfo{booktitle}{Proceedings of the 38th ACM SIGPLAN
  Conference on Programming Language Design and Implementation}},
  \bibinfo{series}{PLDI 2017}, \bibinfo{publisher}{Association for Computing
  Machinery}, \bibinfo{address}{New York, NY, USA}, pp.
  \bibinfo{pages}{495--509}, \doi{10.1145/3062341.3062347}.

\bibitemdeclare{inproceedings}{Pfenning1999}
\bibitem{Pfenning1999}
\bibinfo{author}{Frank \surnamestart Pfenning\surnameend} \&
  \bibinfo{author}{Carsten \surnamestart Sch\"{u}rmann\surnameend}
  (\bibinfo{year}{1999}): \emph{\bibinfo{title}{System Description: Twelf - A
  Meta-Logical Framework for Deductive Systems}}.
\newblock In: {\sl \bibinfo{booktitle}{Proceedings of the 16th International
  Conference on Automated Deduction: Automated Deduction}},
  \bibinfo{series}{CADE-16}, \bibinfo{publisher}{Springer-Verlag},
  \bibinfo{address}{Berlin, Heidelberg}, pp. \bibinfo{pages}{202--206},
  \doi{10.1007/3-540-48660-7_14}.

\bibitemdeclare{article}{Rosu2010}
\bibitem{Rosu2010}
\bibinfo{author}{Grigore \surnamestart Rosu\surnameend} \&
  \bibinfo{author}{Traian~F. \surnamestart
  \c{S}erb\u{a}nu\c{t}\u{a}\surnameend} (\bibinfo{year}{2010}):
  \emph{\bibinfo{title}{{An overview of the K semantic framework}}}.
\newblock {\sl \bibinfo{journal}{The Journal of Logic and Algebraic
  Programming}} \bibinfo{volume}{79}(\bibinfo{number}{6}), pp.
  \bibinfo{pages}{397--434}, \doi{10.1016/j.jlap.2010.03.012}.

\bibitemdeclare{inproceedings}{fab}
\bibitem{fab}
\bibinfo{author}{Gabriel \surnamestart Scherer\surnameend},
  \bibinfo{author}{Max~S. \surnamestart New\surnameend}, \bibinfo{author}{Nick
  \surnamestart Rioux\surnameend} \& \bibinfo{author}{Amal \surnamestart
  Ahmed\surnameend} (\bibinfo{year}{2018}): \emph{\bibinfo{title}{Fabulous
  Interoperability for {ML} and a Linear Language}}.
\newblock In: {\sl \bibinfo{booktitle}{Foundations of Software Science and
  Computation Structures - 21st International Conference, {FOSSACS} 2018, Held
  as Part of the European Joint Conferences on Theory and Practice of Software,
  {ETAPS} 2018, Thessaloniki, Greece, April 14-20, 2018, Proceedings}}, pp.
  \bibinfo{pages}{146--162}, \doi{10.1007/978-3-319-89366-2\_8}.

\bibitemdeclare{inproceedings}{Sewell:2007}
\bibitem{Sewell:2007}
\bibinfo{author}{Peter \surnamestart Sewell\surnameend},
  \bibinfo{author}{Francesco~Zappa \surnamestart Nardelli\surnameend},
  \bibinfo{author}{Scott \surnamestart Owens\surnameend},
  \bibinfo{author}{Gilles \surnamestart Peskine\surnameend},
  \bibinfo{author}{Thomas \surnamestart Ridge\surnameend},
  \bibinfo{author}{Susmit \surnamestart Sarkar\surnameend} \&
  \bibinfo{author}{Rok \surnamestart Strni\v{s}a\surnameend}
  (\bibinfo{year}{2007}): \emph{\bibinfo{title}{Ott: Effective Tool Support for
  the Working Semanticist}}.
\newblock In: {\sl \bibinfo{booktitle}{Proceedings of the 12th ACM SIGPLAN
  International Conference on Functional Programming}}, \bibinfo{series}{ICFP
  '07}, \bibinfo{publisher}{ACM}, \bibinfo{address}{New York, NY, USA}, pp.
  \bibinfo{pages}{1--12}, \doi{10.1145/1291151.1291155}.

\end{thebibliography}

\end{document}